\documentstyle[epsf,rotating]{prhep97}

\newcommand{\lsim}{\raisebox{-0.13cm}{~\shortstack{$<$ \\[-0.07cm] $\sim$}}~}
\newcommand{\gsim}{\raisebox{-0.13cm}{~\shortstack{$>$ \\[-0.07cm] $\sim$}}~}
\newcommand{\tgb}{\mbox{tg}\beta}

\makeatletter
\let\chapter\hid@chapter
\makeatother
\begin{document}

\setcounter{page}{0}

\begin{Large}
\begin{flushright}
CERN-TH/97-323 \\
hep-ph/9711394 \\
November 1997
\end{flushright}
\end{Large}

\vspace*{1cm}

\renewcommand{\thefootnote}{\fnsymbol{footnote}}

\begin{center}
{\Large \bf Higgs Production and Decay at Future Machines\footnote{Contribution
to the proceedings of the {\it International Europhysics Conference on
High-Energy Physics}, 19--26 August 1997, Jersusalem, Israel}}

\vspace*{1cm}

{\Large Michael Spira}

\vspace*{1cm}

{\Large \it CERN, Theory Division, CH-1211 Geneva 23, Switzerland}

\vspace*{2cm}

{\Large \bf Abstract}
\end{center}

\begin{Large}
\noindent
Higgs boson production at future colliders within the Standard Model and its
minimal supersymmetric extension is reviewed. The predictions for decay rates
and production cross sections are presented including all relevant
higher-order corrections.

\vspace*{\fill}

\begin{flushleft}
CERN-TH/97-323 \\
hep-ph/9711394 \\
November 1997
\end{flushleft}
\end{Large}

\thispagestyle{empty}

\input{twocolumn.sty}


\authorrunning{M.\,Spira}
\titlerunning{{\talknumber}: Higgs Production and Decay}
 

\def\talknumber{1805} 

\title{{\talknumber}: Higgs Production and Decay at Future Machines
\vspace*{-0.5cm}}
\author{Michael\,Spira~~~(Michael.Spira@cern.ch)}
\institute{CERN, Theory Division, CH-1211 Geneva 23, Switzerland}

\maketitle
\vspace*{-0.8cm}

\begin{abstract}
Higgs boson production at future colliders within the Standard Model and its
minimal supersymmetric extension is reviewed. The predictions for decay rates
and production cross sections are presented including all relevant
higher-order corrections.
\end{abstract}
\vspace*{-0.9cm}
\section{Introduction}
\vspace*{-0.3cm}
\noindent
\underline{\it Standard Model [SM].}
The SM contains one isospin doublet of Higgs fields, which leads to the
existence of one elementary neutral CP-even Higgs boson $H$ \cite{higgs}. Only
its mass is unknown. The direct search for the Higgs boson at the LEP
experiments excludes Higgs masses below $\sim 77$~GeV \cite{janot}.
Unitarity of scattering amplitudes requires
the introduction of a cut-off $\Lambda$ \cite{trivial}, which imposes an upper
bound on the Higgs mass. For the minimal value $\Lambda
\sim 1$ TeV the upper bound on the Higgs mass is $M_H\lsim 700$~GeV. For
$\Lambda\sim M_{GUT}\sim 10^{15}$ GeV, the Higgs mass has to be smaller than
$\sim 200$ GeV.

\noindent
\underline{\it Minimal Supersymmetric Extension [MSSM].}
Supersymmetry provides a solution to the hierarchy problem of the SM, which
arises for small Higgs masses. The MSSM
requires the introduction of two Higgs doublets, which leads to
five elementary Higgs particles: two neutral CP-even ($h,H$),
one neutral CP-odd ($A$) and two charged ($H^\pm$) Higgs bosons. In
the Higgs sector only two parameters have to be
introduced, which are usually chosen as $\tgb=v_2/v_1$, the ratio of the two
vacuum expectation values of the CP-even Higgs fields, and the pseudoscalar
Higgs mass $M_A$. Radiative corrections to the MSSM Higgs sector are
large, since the leading part grows as the fourth power of the top quark
mass. They increase the upper limit on the light scalar Higgs
mass to $M_h \lsim 130$ GeV \cite{mssmrad}.
\vspace*{-0.5cm}
\section{Standard Model}
\vspace*{-0.3cm}
\subsection{Decay Modes}
\vspace*{-0.2cm}
\noindent
\underline{\it $H\to f\bar f$.}
For $M_H\lsim 140$ GeV the branching ratios of $H\to b\bar b~(\tau^+\tau^-)$
reach values of $\sim 90$\% ($\sim 10$\%).
Above the $t\bar t$ threshold the branching ratio of $H\to t\bar t$
amounts to $\lsim 20$\%. QCD corrections to the $b\bar b, c\bar c$ decays are
large, owing to large logarithmic contributions, which can be absorbed in the
running quark mass $\overline{m}_Q(M_H)$. Far above the $Q\bar Q$ threshold
they are known up to three loops \cite{hbbm,hbb}. The NLO corrections have
been evaluated including the full quark mass dependence \cite{hbbm}.
They are moderate in the threshold region $M_H \gsim 2m_Q$.

\noindent
\underline{\it $H\to W^+W^-, ZZ$.}
The $H\to WW,ZZ$ decay modes dominate for $M_H\gsim 140$ GeV.
Electroweak corrections are small in the intermediate Higgs mass range, while
they enhance the partial widths by about 20\% for $M_H\sim 1$~TeV due to the
self-interaction of the Higgs particle \cite{hvvelw}. For $M_H<2M_{W,Z}$
off-shell decays $H\to W^*W^*, Z^*Z^*$ are important. They lead to $WW$ ($ZZ$)
branching ratios of about 1\% for $M_H\sim 100~(110)$ GeV.

\noindent
\underline{\it $H\to \gamma\gamma$.}
The decay $H\to\gamma\gamma$ is mediated by fermion and $W$-boson
loops, which interfere destructively. The $W$-boson contribution dominates
\cite{hgam}.
The photonic branching ratio reaches a level of $\gsim 10^{-3}$ for Higgs
masses $M_H \lsim 150$ GeV. This decay mode plays a significant r\^ole for the
Higgs search at the LHC for $M_H\lsim 140$ GeV. QCD corrections are
small in the intermediate mass range \cite{gghqcd}.

\begin{figure}

\vspace*{-0.5cm}
\hspace*{-0.5cm}
\begin{turn}{-90}%
\epsfxsize=5cm \epsfbox{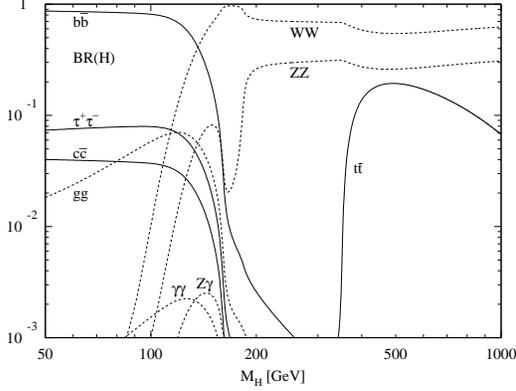}
\end{turn}
\vspace*{-0.3cm}

\caption[]{\label{fg:smbrw} Branching ratios
of the SM Higgs boson as a function of its mass \cite{habil}.}
\vspace*{-1.0cm}
\end{figure}

\noindent
\underline{\it Branching ratios and decay width.}~The~bran\-ching
ratios of the Higgs boson are presented in
Fig.~\ref{fg:smbrw}. For $M_H \lsim 140$ GeV, where the $b\bar b, \tau^+\tau^-,
c\bar c$ and $gg$ decay modes are dominant, the total decay width is very
small, $\Gamma \lsim 10$ MeV. Above this mass value the $WW,ZZ$ decay modes
become
dominant. For $M_H\lsim 2M_Z$ the decay width amounts to $\Gamma\lsim 1$ GeV,
while it reaches $\sim 600$~GeV for $M_H\sim 1$~TeV. Thus the intermediate
Higgs boson is a narrow resonance.
\vspace*{-0.3cm}
\subsection{Higgs Boson Production.}
\vspace*{-0.2cm}
\noindent
\underline{\it $e^+e^-$ colliders.}
\begin{figure}

\vspace*{-3.65cm}
\hspace*{-2.5cm}
\epsfxsize=11.5cm \epsfbox{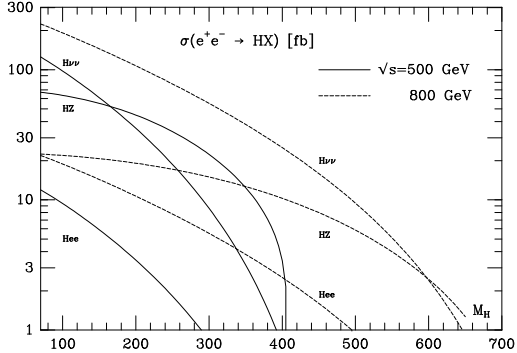}
\vspace*{-0.4cm}

\caption[]{\label{fg:smeepro} Production cross sections of SM Higgs bosons at
future $e^+e^-$ linear colliders \cite{lincoll}.}
\vspace*{-0.75cm}
\end{figure}
At lower energies Higgs bosons are dominantly produced via Higgs-strahlung off
$Z$ bosons, $e^+e^- \to ZH$, while at high energies the $W$ fusion process
$e^+e^- \to \nu_e \bar \nu_e H$ dominates \cite{lincoll}.
Electroweak
corrections to the Higgs-strahlung process are moderate \cite{hzelw}. The
production cross sections at
future $e^+e^-$ colliders are shown in Fig.~\ref{fg:smeepro} for c.m.~energies
of 500 and 800 GeV. They range between 1 and 300 fb in the relevant
mass range and provide clean signatures for the Higgs boson. The
angular distribution of Higgs-strahlung is sensitive to
the spin and parity of the Higgs particle \cite{lincoll}.

\noindent
\underline{\it LHC.}
\begin{figure}

\vspace*{-0.55cm}
\hspace*{-0.45cm}
\begin{turn}{-90}%
\epsfxsize=5cm \epsfbox{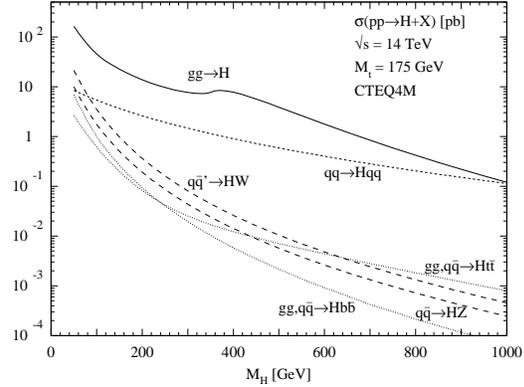}
\end{turn}
\vspace*{-0.3cm}

\caption{\label{fg:smlhcpro} Higgs production cross sections at the LHC for
the various mechanisms as a function of the Higgs mass \cite{habil}.}
\vspace*{-0.75cm}
\end{figure}
Higgs boson production at the LHC is dominated by the gluon fusion mechanism
$gg\to H$, which is mediated by top and bottom triangle loops
\cite{gghqcd,glufus}.
This can be inferred from Fig.~\ref{fg:smlhcpro}, which presents all relevant
Higgs production cross sections as a function of the Higgs mass.
The two-loop QCD corrections increase the production cross section by 60--90\%,
so that they can no longer be neglected \cite{gghqcd}.
[The QCD corrections to most of
the background processes at the LHC are also known.]
In spite of the large size of the QCD corrections the
scale dependence is reduced significantly, thus rendering the NLO result
reliable.

For large Higgs boson masses the vector boson fusion mechanism $WW,ZZ \to H$
becomes competitive, while for intermediate Higgs masses it is about an order
of magnitude smaller than gluon fusion \cite{wwh}. The QCD corrections are
small and thus negligible \cite{wwhqcd}.

Higgs-strahlung $W^*/Z^* \to H W^*/Z^*$ plays a r\^ole only
for $M_H\lsim 100$ GeV. The QCD corrections are moderate,
so that this process is calculated with reliable accuracy \cite{whqcd}.

Higgs bremsstrahlung off top quarks, $gg,q\bar q \to Ht\bar t$, is
sizeable for $M_H\lsim 100$ GeV \cite{kunszt}. The QCD
corrections to this process are unknown, so that the cross section is uncertain
within a factor of $\sim 2$.
\vspace*{-0.5cm}
\section{MSSM}
\vspace*{-0.3cm}
\subsection{Decay Modes}
\vspace*{-0.2cm}
Typical examples of the branching ratios and decay widths of the MSSM Higgs
bosons can be found in Refs.~\cite{habil,lincoll}.

\noindent
\underline{\it $\phi\to f\bar f$.}
For large $\tgb$ the decay modes $h,H,A\to b\bar b, \tau^+\tau^-$
dominate the neutral Higgs decay modes, while for small $\tgb$ they
are important for $M_{h,H,A}\lsim 150$ GeV. The dominant decay modes of
charged Higgs particles are $H^+\to \tau^+\nu_\tau, t\bar b$.
The QCD corrections reduce the partial decay widths into $b,c$ quarks by
50\%--75\% as a result of the running quark masses, while they are moderate
for decays into top quarks \cite{hbbm,hbb}.

\noindent
\underline{\it Decays into Higgs and gauge bosons.}
The decay modes $H\to hh,AA,ZA$ and $A\to Zh$ are important for small $\tgb$
below the $t\bar t$ threshold. Similarly the decays $H^+\to WA,Wh$ are sizeable
for small $\tgb$ and $M_{H^+}< m_t+m_b$. The dominant higher-order corrections
can be absorbed into the couplings and masses of the Higgs sector. Below the
corresponding thresholds decays into off-shell Higgs and gauge bosons turn out
to be important especially for small $\tgb$ \cite{offshell}. The decays
$h,H\to WW,ZZ$ are
suppressed by kinematics and, in general, by the SUSY couplings and are thus
less important in the
MSSM. The decay $h\to \gamma\gamma$ is only relevant for
the LHC in the decoupling limit, where the light scalar Higgs boson
$h$ has similar properties to those of the SM Higgs particle.

\noindent
\underline{\it Decays into SUSY particles.}~Higgs~decays~in\-to
charginos, neutralinos and third-generation sfermions can become
important, once they are kinematically allowed \cite{SUSY}. Thus
they could
complicate the Higgs search at the LHC, since the decay into the LSP will be
invisible.

\vspace*{-0.5cm}
\subsection{Higgs Boson Production}
\vspace*{-0.2cm}
\noindent
\underline{\it $e^+e^-$ colliders.}
\begin{figure}

\vspace*{-3.3cm}
\hspace*{-2.2cm}
\epsfxsize=11.5cm \epsfbox{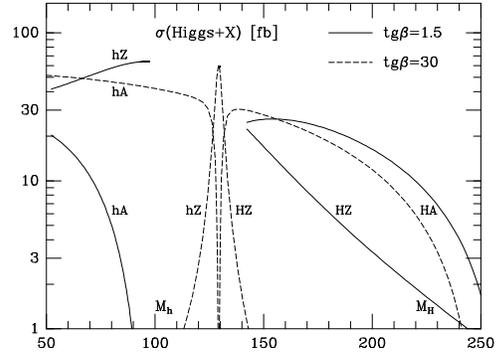}
\vspace*{-1.8cm}

\caption{\label{fg:mssmeepro} Production cross sections of MSSM Higgs bosons at
future $e^+e^-$ linear colliders \cite{lincoll}.}
\vspace*{-0.7cm}
\end{figure}
Neutral MSSM Higgs bosons will be produced dominantly via $e^+e^-\to Z+h/H, A +
h/H$ and $W$ boson fusion $e^+e^-\to \nu_e\bar\nu_e + h/H$ at
future $e^+e^-$ colliders. The size of the individual cross sections depends
strongly
on $\tgb$, but the sum is always of the order of the SM cross section. Typical
examples are presented in Fig.~\ref{fg:mssmeepro} for
$\sqrt{s} = 500$ GeV, ranging between 1 and 100 fb \cite{lincoll}.

Charged Higgs bosons can be produced via pair production $e^+e^-\to H^+H^-$ or
through top quark decays $e^+e^-\to t\bar t \to H^+ b \bar t$. They are in
general detectable, with masses up to half the energy of the $e^+e^-$
collider \cite{lincoll}.

\noindent
\underline{\it LHC.}
\begin{figure}

\vspace*{0.3cm}
\hspace*{-0.45cm}
\begin{turn}{-90}%
\epsfxsize=5cm \epsfbox{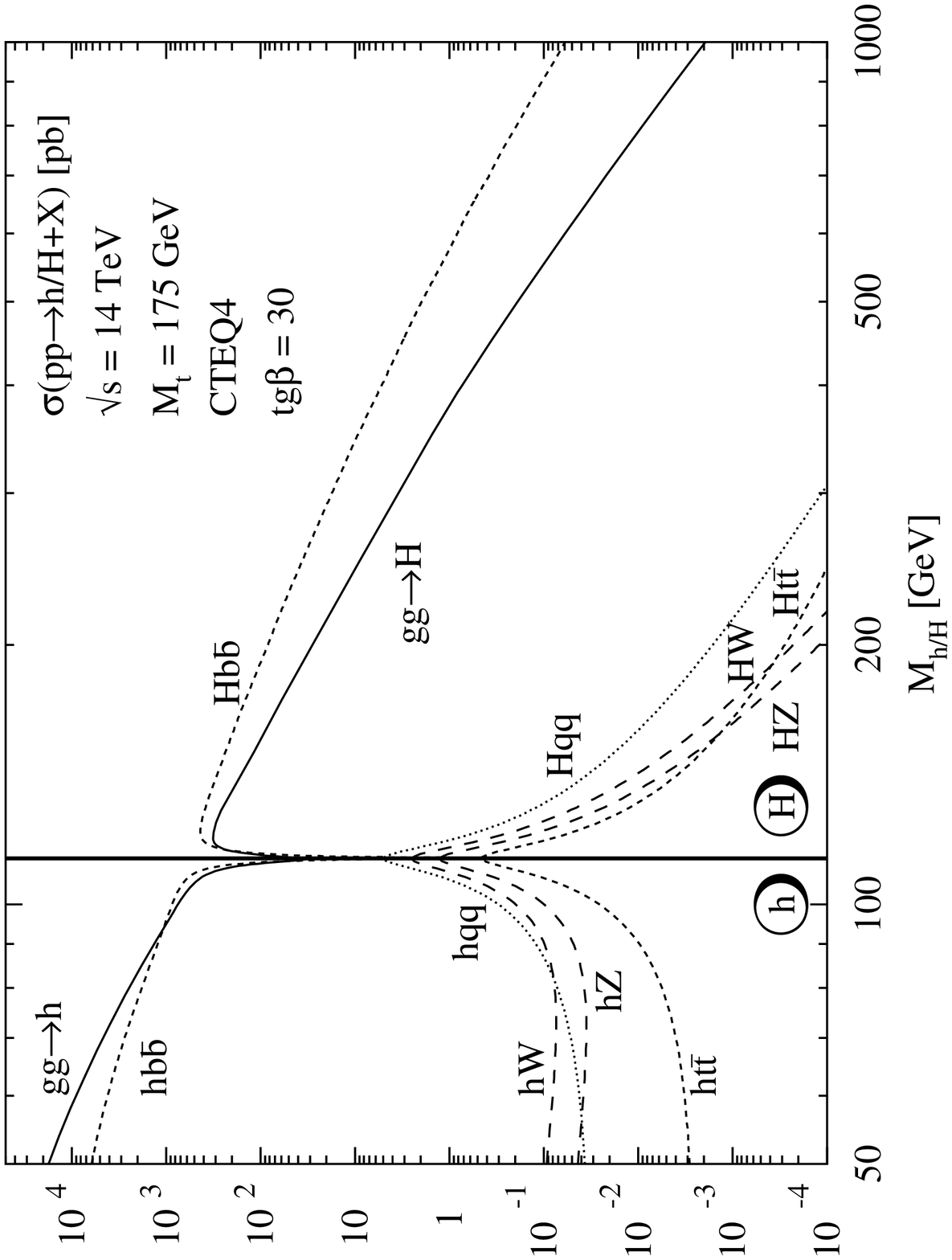}
\end{turn}
\vspace*{-0.3cm}

\caption{\label{fg:mssmlhcpro} Scalar Higgs production cross sections at the
LHC for the various mechanisms as a function of the Higgs mass for $\tgb=30$
\cite{habil}.}
\vspace*{-0.6cm}
\end{figure}
As can be inferred from Fig.~\ref{fg:mssmlhcpro}, neutral MSSM Higgs bosons
are dominantly produced via gluon fusion, $gg\to h,H,A$, which is mediated by
top and bottom quark loops \cite{gghqcd,glufus}. Only for squark masses
below $\sim 400$ GeV can the squark loop contributions become significant
\cite{squloop}. The
two-loop QCD corrections increase the production cross sections by 10\%--100\%
and thus cannot be neglected \cite{gghqcd}.

For large $\tgb$ Higgs bremsstrahlung off $b$ quarks, $gg,q\bar q \to
b\bar b +h/H/A$, dominates in a large part of the relevant Higgs mass
ranges \cite{kunszt,pphbb}. The QCD corrections are unknown.

Vector boson fusion $WW/ZZ\to h/H$ and Higgs-strahlung,
$W^*/Z^*\to W/Z + h/H$, are suppressed in most of the parameter space by
SUSY couplings compared with the SM and thus less important.

\vspace*{-0.5cm}
%

\end{document}